\documentclass[journal]{IEEEtran}
\usepackage{cite}
\usepackage[pdftex]{graphicx}
\usepackage{epstopdf} %included
\graphicspath{{./figures/}{./pictures/}}
\usepackage{array}
\usepackage[small]{caption}
\usepackage{verbatim}
\usepackage{graphicx}% remove the 'demo' option in the real document
\usepackage{color} % to color the comments

\usepackage{graphicx}% http://ctan.org/pkg/graphicx
\usepackage{stfloats}
\usepackage[cmex10]{amsmath}
\usepackage{amsfonts}
\usepackage{amsthm,amssymb,url,amsmath}
\usepackage[tight,footnotesize]{subfigure}
\usepackage{url}
\usepackage{booktabs}
\usepackage[utf8]{inputenc}
\usepackage{mathrsfs}
\usepackage{xspace}
\usepackage{xfrac}
\usepackage{multirow}
\usepackage{balance}

\RequirePackage[utf8]{inputenc}
\usepackage{booktabs}
\usepackage{mathtools}
\usepackage{algorithmic}
\usepackage{algorithm}

\makeatletter
\newcommand{\removelatexerror}{\let\@latex@error\@gobble}
\makeatother

\usepackage{tikz}
\usepackage{pgfplots}
\tikzset{every picture/.append style={font=\scriptsize}}
\usetikzlibrary{external,calc} %jc
\tikzset{%
  >=latex, % option for nice arrows
  inner sep=0pt,%
  outer sep=2pt,%
  mark coordinate/.style={inner sep=0pt,outer sep=0pt,minimum size=3pt,
    fill=black,circle}%
}

\newcommand{\R}{{\mathbb R}}

\newcommand{\T}{\text{T}}

\newcommand{\dd}{{\mathrm{d}}}

\newcommand{\0}{\boldsymbol{0}}

\newcommand{\G}{\text{G}}

\newcommand{\g}{\mathfrak{g}}
\newcommand{\SE}{\text{SE}}

\newcommand{\SO}{\text{SO}}

\newcommand{\I}{\text{I}}

\renewcommand{\tt}{t}

\newcommand{\vMises}{v\kern.01emM\xspace}

\newcolumntype{C}[1]{>{\centering}m{#1}}

\DeclareMathOperator{\ad}{ad}
\DeclareMathOperator{\Ad}{Ad}

\newcommand{\im}[1]{{\textcolor{blue}{{[im:#1]}}}}

\newcommand{\jc}[1]{{\textcolor{red}{{[jc:#1]}}}}

\begin{document}

% *** PAPER TITLE ***
%
\title{%Multitarget tracking employing 
Mixture Reduction on Matrix Lie Groups \vspace{-2mm}
}

\author{Josip~Ćesić,~\IEEEmembership{Member,~IEEE,}
        Ivan~Marković,~\IEEEmembership{Member,~IEEE,}
        and~Ivan~Petrović,~\IEEEmembership{Member,~IEEE}% <-this % stops a space
$^{\ast}$\vspace{-6mm}
%\thanks{}
%\thanks{
%This work has been supported by European Community's Seventh Framework Programme
%under Grant agreement No.~285939 (ACROSS).
%}
\thanks{
$^{\ast}$J. Ćesić, I. Marković and I. Petrović are with the University of Zagreb 
Faculty of Electrical Engineering and Computing,
%Department of Control and Computer Engineering,
Unska 3, 10000 Zagreb, Croatia.
{E-mail: \{name.surname@fer.hr\}}

This work has been supported from the Unity Through
Knowledge Fund (no. 24/15) under the project Cooperative Cloud
based Simultaneous Localization and Mapping in Dynamic
Environments (cloudSLAM) and the European Union’s
Horizon 2020 research and innovation programme under
grant agreement No 688117 (SafeLog).
}
}

\maketitle

\begin{abstract}
Many physical systems evolve on matrix Lie groups and mixture filtering designed for such manifolds represent an
inevitable tool for challenging estimation problems.
However, mixture filtering faces the issue of a constantly growing number of components, hence require
appropriate mixture reduction techniques.
In this letter we propose a mixture reduction approach for distributions on matrix Lie groups, called the
concentrated Gaussian distributions (CGDs).
This entails appropriate reparametrization of CGD parameters to compute the KL divergence, pick and merge the mixture
components.
Furthermore, we also introduce a multitarget tracking filter on Lie groups as a mixture filtering study example for the
proposed reduction method. 
In particular, we implemented the probability hypothesis density filter on matrix Lie groups.
We validate the filter performance using the optimal subpattern assignment metric on a synthetic dataset consisting of
$100$ randomly generated multitarget scenarios.
\end{abstract}


\IEEEpeerreviewmaketitle

%\section{Introduction}
%\label{sec:intro}\noindent

\vspace{-3mm}
\section{Introduction}
\label{sec:intro}%\subsubsection{General intro on Mixture filtering}
\noindent
\IEEEPARstart{M}{any} statistical and engineering problems require modeling of complex multi-modal data, wherein mixture distributions became an inevitable tool \cite{Alspach1972,Chen2000},
primarily in traditional application domains like
radar and sonar tracking \cite{Blackman1986},
and later in different modern fields such as
computer vision \cite{Stauffer2000},
speech recognition \cite{Goldberger2005} or
multimedia processing \cite{Nikseresht2008}.
Approaches relying on mixture distributions often face the problem of large or an ever increasing number of mixture components, hence the growth of components must be controlled by approximating the original mixture with the mixture of a reduced size \cite{Salmond1989, Runnalls2007, Bukal2014a}.
%\cite{Alspach1972, Salmond1989, Salmond2009, Runnalls2007, Vo2006, Vo2007, Mahler2007, Vo2009, Reuter2014a,Williams2015a,Markovic2015}.
For example, in the case of multitarget tracking applications,
by employing conventional Gaussian mixture based filters \cite{Vo2006,Vo2007}, 
during the recursion process the number of components inevitably increases.
This appears firstly due to appearance of newly birthed or spawned components, 
and secondly, due to inclusion of multiple measurements, which results in
geometrical increase in the number of components.
%the number of components, and hence the reduction procedures are necessary.

Another important aspect of estimation is the state
space geometry, hence many works have been dedicated
to appropriate uncertainty modeling and estimation techniques
for a wide range of applications 
\cite{Lui2012,Loianno2016,Cesic2016b,Barrau2015}, motivated by theoretical and implementation difficulties caused by
treating a constrained problem naively with Euclidean tools.
%On the contrary, accounting for the geometry of the state space leads to well-posed problems which can increase the performance \cite{Bourmaud2015}.
%due to the fact that treating a constrained problem naively by Euclidean space tools may cause theoretical and implementation difficulties \cite{Lui2012,Hertzberg2013,Bourmaud2013,Bourmaud2015, Bourmaud2015a,Bourmaud2016,Berger2015,Loianno2016,Cesic2016c}.
For example, Lie groups are natural
ambient (state) spaces for description of the dynamics
of rigid body mechanical systems.
In \cite{Long2012} it has been observed that the distribution of the pose of a differential drive mobile robot is not
a Gaussian distribution in Cartesian coordinates, but rather a distribution on the special Euclidean group $\SE(2)$.
Similarly, \cite{Barfoot2014} discussed the uncertainty association with 3D pose employing the $\SE(3)$ group.
Furthermore, attitude estimation arises naturally on the $\SO(3)$ group \cite{Barrau2015}.
In \cite{Zhang2016} a feedback particle filter on matrix Lie groups was proposed, while
in \cite{Bourmaud2013,Bourmaud2015} authors proposed an extended Kalman filter on matrix Lie
groups (LG-EKF), building the theory upon the concentrated Gaussian distribution (CGD) on matrix Lie groups \cite{Wolfe2011}.
%Ito2000 - navigational and guidance systems, radar tracking, sonar ranging, and satellite and airplane orbit determination
%Bukal2014 - image and multimedia indexing [8,9], speech segmentation [10], and it is an indispensable part of any tracking system with mixtures of Gaussian [11–13] or von Mises distributions [3].

%Considering the application-wise perspective of the Lie group ne
%Rigid body motion generally does not occur in euclidean space, but rather arise on curved geometries often called manifolds.

%\subsubsection{Related work on mixture filtering on Lie groups}\noindent
%Estimation of complex data by mixture distributions may lead to models with large or, in applications like multitarget tracking (MTT), ever increasing number of components \cite{Bukal2014a,Vo2006}.

In this letter we address finite mixtures of distributions on matrix Lie groups.
We propose a novel approach to CGD mixture reduction, which required finding solutions for computing Kullback-Leibler
divergence of CGD components and CGD component merging.
Furthermore, since previous methods require choosing the appropriate tangent space, we also provide an extensive analysis
on the choice thereof.
As a study example, we use the proposed reduction method in a multitarget tracking scenario.
We introduce the probability hypothesis density filter (PHD) on matrix Lie groups with approximation based on a finite
mixture of CGDs.

\vspace{-1mm}
\section{Mathematical Preliminaries}
\label{sec:preliminaries}
\noindent
%We first present the theoretical preliminaries addressing Lie
%groups and uncertainty definition in the form of the CGD.
We now introduce theoretical preliminaries concerning Lie groups; however, for a more rigorous introduction
the reader is directed to \cite{Chirikjian2012}.
A Lie group $\G$ is a group which has the structure of a smooth manifold; moreover, 
a tangent space $T_X(\G)$ is associated to $X\in\G$ such that 
the tangent space placed at the group identity,
called Lie algebra $\g$, is transferred by applying corresponding action to $X$. 
%, called the Lie algebra of $\G$,
%is associated to $X\in\G$ through a 
%each point $X\in\G$ has an
%associated tangent space $T_X(\G)$, called the Lie algebra of $\G$, which almost completely captures a curved object
%like $\G$.
%This linear tangent space is usually placed at the group identity, 
%%simply because it is the most natural one to consider, 
%and is called the Lie algebra of $\G$, which we denote by $\g$ \cite{Selig2005}.
%The Lie algebra $\g$, which is of the same dimension as $\G$, has an extra structure; a binary operation $[\cdot,\cdot]$
%called the Lie bracket, which reflects the non-commutative content of the group operation.
%Furthermore, if the group $\G$ is a matrix Lie group, then $\G \subset \R^{n\times n}$ and group operations are simply
%matrix multiplication and inversion.
In this paper we are interested in matrix Lie groups which are usually the ones considered in engineering and physical
sciences.
%Moreover, the theorem \cite{Ado1947} says that every Lie algebra is isomorphic to a matrix Lie algebra, thus we will simply
%say `Lie algebra' rather than `matrix Lie algebra'.

The Lie algebra $\g\subset \R^{n \times n}$ associated to a $p$-dimensional matrix Lie group $\G \subset \R^{n \times
n}$ is a $p$-dimensional vector space.
%; particularly, an open neighbourhood around $\0^p$ in the tangent space of $\G$ at
%the identity $\I^n$.
% defined by a basis consisting of $p$ real matrices $E_r$, $r=1,\dots,p$ \cite{Park2010a}. 
The matrix exponential $\exp_{\G}$ and matrix logarithm $\log_{\G}$ establish a local diffeomorphism between the two
% between $\G$ and $\g$ as
%
\begin{align}
  \exp_{\G} : \g \rightarrow \G \text{ \ and \ }
  \log_{\G} : \G \rightarrow \g.
\end{align}
Furthermore, a natural relation exists between $\g$ and the Euclidean space $\R^p$ given through a linear isomorphism
\begin{align}
  [\cdot]^{\vee}_{\G}  : \g \rightarrow \R^p \text{ \ and \ }
  [\cdot]^{\wedge}_{\G}  : \R^p \rightarrow \g.
\end{align}
For $x\in\R^p$ and $X\in\G$ we use the following notation \cite{Bourmaud2016}
\begin{align}
  \exp_{\G}^{\wedge}(x) = \exp_{\G}( [x]^{\wedge}_{\G}) \text{ \ and \ }
  \log^{\vee}_{\G}(X) = [\log_{\G} (X)]^{\vee}_{\G} \,.
\end{align}
%
%where $x\in\R^p$ and $X\in\G$.

Lie groups are generally non-commutative, i.e., $XY \neq YX$.
However, the non-commutativity can be captured by the so-called adjoint representation of $\G$ on $\g$
\cite{Bourmaud2015a}
\begin{equation}\label{eq:lg_switcharoo}
  X\exp^{\wedge}_{\G}(y) = \exp^{\wedge}_{\G}(\Ad_{\G}(X)y)X,
\end{equation}
which can be seen as a way of representing the elements of the group as a linear transformation of the group’s
algebra.
The adjoint representation of $\g$, $\ad_{\G}$, is in fact the differential of $\Ad_{\G}$
at the identity.
Another important result for working with Lie group elements is the Baker-Campbell-Hausdorff (BCH) formula, which enables
representing the product of Lie group members as a sum in the Lie algebra.
We will use the following BCH formulae \cite{Miller1972,Bourmaud2015a}
\begin{align}
\label{eq:bch2}
 &  \log^{\vee}_{\G}(\exp_{\G}^{\wedge}(x)\exp_{\G}^{\wedge}(y)) = y + \varphi_{\G}(y) x + O(||y||^2), \\ 
\label{eq:bch3}
 &  \log^{\vee}_{\G}(\exp_{\G}^{\wedge}(x+y)\exp_{\G}^{\wedge}(-x)) = \Phi_{\G}(x) y + O(||y||^2), 
\end{align}
where $\varphi_{\G}(y) = \sum_{n=0}^{\infty} \frac{B_n\ad_{\G}(y)^n}{n!}$, $B_n$ are Bernoulli numbers,
and $\Phi_{\G}(x)=\varphi_{\G}(x)^{-1}$.
For many common groups used in engineering and physical sciences closed form expressions for $\varphi_{\G}(\cdot)$ and $\Phi_{\G}(\cdot)$ can be found \cite{Barfoot2014,Bourmaud2015a}; otherwise, a truncated series expansion is used.

\begin{comment}
%\subsection{Concentrated Gaussian distribution}
%\noindent
%To make use of EKF on Lie groups, we need to establish first a notion of a Gaussian distribution on Lie groups.
A distribution on a Lie group for which almost all the mass of the
distribution is concentrated in a small neighborhood around the mean, can be expressed in the Lie algebra
\cite{Wang2008,Barfoot2014}, and this concept is called a \emph{concentrated Gaussian distribution}.
Let $X\in\G$ be a random variable that follows a CGD with mean $\mu$ and covariance
$\Sigma$ as
%
\begin{equation}\label{eq:cgd_parametric}
\vspace{-0.0mm}
  X = \exp^{\wedge}_{\G}(\xi)\mu, \quad X \sim \mathcal{G}(\mu,\Sigma),
\vspace{-0.0mm}
\end{equation}
%
where $\xi \sim \mathcal{N}_{\R^p}(\0^{p},\Sigma)$ is a zero-mean Gaussian distribution with covariance
$P\subset\R^{p\times p}$ defined in the Lie algebra, i.e., the Euclidean space $\R^p$.
\end{comment}

\begin{comment}
\begin{figure}[t]
\centering
  \begin{picture}(100,65)
	\put(-50,0){\includegraphics[width=0.8\columnwidth]{CGD_new.pdf}}
	\put(-45,0){$\G$}
%	\put(-70,23){$\g$}
	\put(-72,48){$w^a, \mathcal{G}(\mu^a, \Sigma^a)$}
	\put(-17,20){$\mu^a$}
%	\put(28,58){$\g$}
	\put(-3,57){$w^b, \mathcal{G}(\mu^b, \Sigma^b)$}
	\put(55,37){$\mu^b$}
%	\put(174,9){$\g$}
	\put(117,45){$w^c, \mathcal{G}(\mu^c, \Sigma^c)$}
	\put(122,17){$\mu^c$}
  \end{picture}
  \caption{Mixture of the concentrated Gaussian distribution components.}
  \vspace{-4mm}
\label{fig:CGD_mixture}
\end{figure}
\end{comment}

\vspace{-2mm}
\subsection{Concentrated Gaussian distribution}\noindent
Herein we introduce the concept of the concentrated Gaussian distribution which is used to define random variables on matrix Lie group.
%The concept of the concentrated Gaussian distribution is necessary for defining random variables on matrix Lie groups.
A random variable $X\in\G$ has a CGD with the mean $\mu$ and covariance $\Sigma$, i.e., $X \sim \mathcal{G}(X;
\mu, \Sigma)$, if 
\begin{equation}\label{eq:cgd}
  X = \exp_{\G}^{\wedge}( {\xi} )\mu\,,
\end{equation}
where $\mu\in\G$, and $\xi \sim \mathcal{N}(\xi; \0_{p\times 1}, \Sigma)$ is a zero-mean `classical' Gaussian random variable
 with the covariance $\Sigma\subset\R^{p\times p}$ \cite{Barfoot2014,Bourmaud2015}.
Note that in this way, we are directly defining the CGD covariance in the pertaining Lie algebra $\g$, while the
mean is defined on the group $\G$.

Given that, the previous definition \eqref{eq:cgd} then induces a pdf of $X$ over $\G$ as follows
\cite{Bourmaud2015,Barfoot2014}
\begin{align}
1 &= \int_{\R^p} \frac{1}{\sqrt{(2\pi)^p |\Sigma|}}\exp_{\G}^{\wedge}{\Big(-\frac12 || \xi ||_{\Sigma}^2\Big)}\dd\xi\nonumber\\
  &= \int_{\G} \beta \exp_{\G}^{\wedge}\Big({-\dfrac{1}{2} ||\log_{\G}^{\vee}(X\mu^{-1})||_{\Sigma}^2}\Big) \dd X\label{eq:cgd_pdf}
\end{align}
where $||x||_{\Sigma}^2 = x^{\mathrm{T}} \Sigma^{-1} x$.
Therein the change of coordinates $\xi =
\log_{\G}^{\vee}(X\mu^{-1})$, with the pertaining differentials $\dd X = | \Phi(\xi)|\dd\xi$, resulted with the CGD
normalizing constant 
\begin{equation}\label{eq:beta}
  \beta = 1 / \sqrt{(2\pi)^p |\Phi(\log_{\G}^{\vee}(X\mu^{-1})) \Sigma \Phi(\log_{\G}^{\vee}(X\mu^{-1}))^{\mathrm{T}}|}.
\end{equation}
Note that this change of variables is valid if all eigenvalues of $\Sigma$ are small, i.e., almost all the mass of the
distribution is concentrated in a small neighborhood around the mean value \cite{Bourmaud2015}.
The pdf over $X$ is now fully determined by \eqref{eq:cgd_pdf} and \eqref{eq:beta}.

\vspace{-2mm}
\section{CGD mixture reduction}
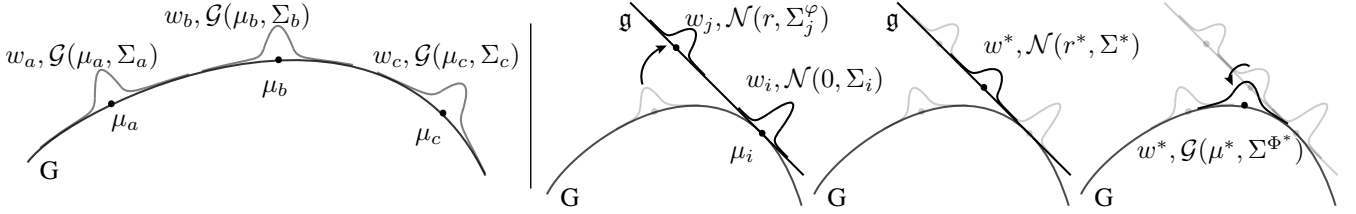
\begin{figure*}[!t]
\centering
\begin{picture}(100,78)
\put(-200,0){\includegraphics[width=2\columnwidth]{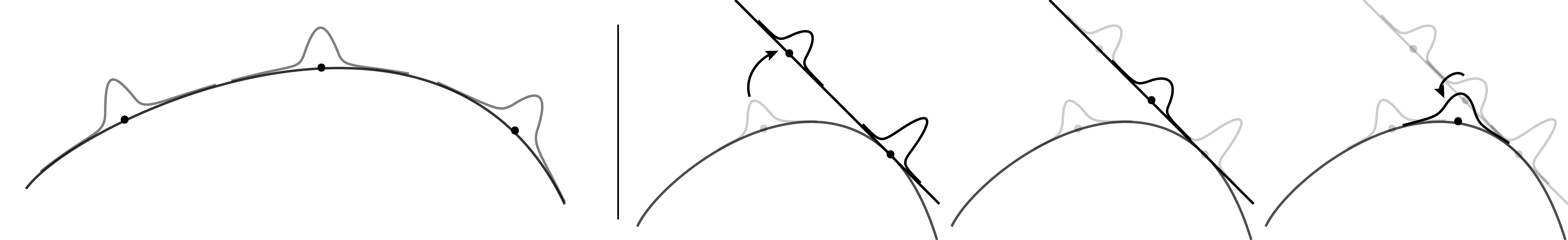}}
	\put(-187,10){$\G$}
	\put(-200,54){$w_a, \mathcal{G}(\mu_a, \Sigma_a)$}
	\put(-160,30){$\mu_a$}
	\put(-140,69){$w_b, \mathcal{G}(\mu_b, \Sigma_b)$}
	\put(-103,43){$\mu_b$}
	\put(-61,54){$w_c, \mathcal{G}(\mu_c, \Sigma_c)$}
	\put(-45,25){$\mu_c$}
%%%%%%%%%%%%%%%%%%%%%%%%%
%\put(-195,0){$\G$}
%\put(-190,55){$w^j, \mathcal{G}(\mu^j, P^j)$}
%\put(-155,28){$\mu^j$}
%\put(-122,48){$w^i, \mathcal{G}(\mu^i, P^i)$}
%\put(-122,20){$\mu^i$}
%%%%%%%%%%%%%%%%%%%%%%%%%
\put(10,0){$\G$}
\put(32,68){$\g$}
\put(57,68){$w_j, \mathcal{N}(r, \Sigma_j^\varphi)$}
\put(80,45){$w_i, \mathcal{N}(0, \Sigma_i)$}
\put(74,18){$\mu_i$}
%%%%%%%%%%%%%%%%%%%%%%%%%
\put(111,0){$\G$}
\put(133,68){$\g$}
\put(170,58){$w^*, \mathcal{N}(r^*, \Sigma^{*})$}
%%%%%%%%%%%%%%%%%%%%%%%%%
\put(212,0){$\G$}
\put(228,18){$w^*, \mathcal{G}(\mu^*, \Sigma^{\Phi^*})$}
\end{picture}
\vspace{-1.0mm}
\caption{
Illustration of a finite mixture of CGDs (left) and the component merging procedure (right).}
\label{fig:illustration}
\vspace{-3mm}
\end{figure*}

\noindent
\iffalse
Approaches relying on density mixtures often face the problem of the increasing number of mixture components
\cite{Alspach1972, Salmond1989, Salmond2009, Runnalls2007, Vo2006, Vo2007, Mahler2007, Vo2009, Reuter2014a,
Williams2015a,Markovic2015}.
For example, in the case of multitarget filters during the recursion process the number of components inevitably increases: first, due to inclusion of newly birthed or spawned components, and second, due to the correction step.
Namely, correcting the predicted mixture by multiple measurements results in a geometrical increase in the number of components, and hence the reduction procedures are necessary.
\fi
With the theoretical preliminaries setup, we continue with mixture reduction on matrix
Lie groups.
A finite mixture of our present interest is given as the weighted sum of CGDs
\begin{align}\label{eq:cgd_sum}
\sum_{i=1}^N w_i \mathcal{G}(X;\mu_i, \Sigma_i)\,,
\end{align}
where $w_i$ are component weights and $N$ is the total number of mixture components.
An illustration of \eqref{eq:cgd_sum} is given in Fig.~\ref{fig:illustration}.
Component reduction procedures typically require three building blocks: (i) component distance measure, (ii) component
picking algorithm, and (iii) component merging.
While various solutions exist for `classical' Gaussian mixtures
\cite{Salmond1989,Runnalls2007,Bukal2014a}, questions
remain on how to approach the component number reduction for CGD mixtures on matrix Lie groups.
Therefore, first we focus on the the fundamental question of how to measure the distance between two CGD components.
\vspace{-2mm}
\subsection{Component distance measure}\label{sec:component_distance_measure}\noindent
Our aim is to use a standard information-theoretic measure
between two CGD components and we propose to use the Kullback-Leibler (KL) divergence \cite{Kullback1997}.
Let $\mathcal{G}_i = \mathcal{G}(X;\mu_i, \Sigma_i)$ and 
$\mathcal{G}_j = \mathcal{G}(X;\mu_j, \Sigma_j)$ be two mixture components with $p_i(X)$ and $p_j(X)$ as their respective
pdfs.
Since there is nothing intrinsic in the definition of KL divergence that requires the underlying space to be Euclidean, by
definition 
% and employing the change of coordinates as in \eqref{eq:pdf_change}, it follows that
%
\begin{align}\label{eq:ss_kl_div}
 \mathcal{D}_{\mathrm{KL}}(\mathcal{G}_i,\mathcal{G}_j)
 &  = \int_{\G} p_i(X) \log \bigg( \dfrac{p_i(X)}{p_j(X)} \bigg) \dd X \,.
\end{align}
In order to evaluate the integral \eqref{eq:ss_kl_div}, we need to employ the change of coordinates as in
\eqref{eq:cgd_pdf}, but this time from the direction of the group $\G$, i.e., from $X\in\G$ to $\xi \in \R^p$.
Note that in \eqref{eq:cgd_pdf} the change evolved around the distribution mean $\mu$; however, since in
\eqref{eq:ss_kl_div} generally $\mu_i \neq \mu_j$, we cannot apply the same approach.
Hence, before evaluating \eqref{eq:ss_kl_div}, we first discuss how to change the coordinates on the level of a
single distribution.

Let $\mathcal{G}(X;\mu,\Sigma)$ be a CGD,
and if we change the coordinates using
$X = \exp^{\wedge}_{\G}(\xi)\mu_{\tt}$, $\mu_{\tt} \in \G$, where $\mu_{\tt} \neq \mu$,
we get
\begin{align}\label{eq:pdf_change}
1 & = \int_{\G} \beta \exp_{\G}^{\wedge} \Big({-\dfrac{1}{2} ||\log_{\G}^{\vee}(X\mu^{-1})||_{\Sigma}^2}\Big) \dd X \\
\nonumber & \stackrel{\text{CoC}}{\approx} \int_{\R^p} \eta \exp^{\wedge}_{\G} 
\Big({-\dfrac{1}{2} ||\log_{\G}^{\vee}(\exp^{\wedge}_{\G}(\xi) \mu_{\tt}\mu^{-1})||_{\Sigma}^2} \Big) \dd \xi \\ % r_{\tt} = \mu \mu_{\tt}^{-1}
\nonumber & \stackrel{\text{\eqref{eq:bch3}}}{\approx} \int_{\R^p} \eta \exp^{\wedge}_{\G}
\Big({-\dfrac{1}{2} 
||\Phi_{\G}(r_{\tt})(\xi-r_{\tt})||_{\Sigma}^2} \Big) \dd \xi \\
\nonumber & = \int_{\R^p} \eta \exp^{\wedge}_{\G} 
\Big({-\dfrac{1}{2}
||\xi-r_{\tt}||^2_{\varphi_{\G}(r_{\tt}) \Sigma  \varphi_{\G}^\T(r_{\tt})} } \Big) \dd \xi \,,
\end{align}
where $r_{\tt} = \log^{\wedge}_{\G}(\mu \mu_{\tt}^{-1})$, $\eta$ approximately evaluates to
\begin{align}\nonumber
\eta = \beta |\Phi(\xi)|
& =
\dfrac{|\Phi(\xi)|}{\sqrt{(2\pi)^p |\Sigma|} \cdot |\Phi(\log_{\G}^{\vee}(\exp^{\wedge}_{\G}(\xi)\mu_{\tt}\mu^{-1}))|} \\ 
& \approx 
\dfrac{1}{\sqrt{(2\pi)^p |\varphi_{\G}(r_{\tt})  \Sigma \varphi_{\G}(r_{\tt})^{\mathrm{T}}|}} \,,
%1 / \sqrt{(2\pi)^p |\varphi_{\G}(r_{\tt})  \Sigma \varphi_{\G}(r_{\tt})^{\mathrm{T}}|} 
\end{align}
and we obtain $\xi \sim \mathcal{N}(\xi; r_{\tt},\varphi_{\G}(r_{\tt})\Sigma \varphi_{\G}^\T(r_{\tt}))$.

\vspace{1mm}
\noindent  \textbf{Remark 1.}
\textit{Covariance of a CGD represents the uncertainty relevant only to the tangent space of its own mean.
In \cite{Bourmaud2015a} authors studied how the covariance changes if looked at from the perspective of a value
which is different than the distribution mean.
They dubbed this procedure `distribution unfolding'.
For example, if we unfold $\mathcal{G}(X;\mu, \Sigma)$ around an arbitrary $\mu_t \in \G$, using
\eqref{eq:bch2} and following \cite{Bourmaud2015a} we get
\begin{align} %\label{eq:cgd_unfolding_ij}
  \xi_{\tt} &= \log^{\vee}_{\G} \big(\exp^{\wedge}_{\G}(\xi) \mu\mu_{\tt}^{-1} \big) \nonumber\\ 
  &\hspace{-0mm} 
  \approx \log^{\vee}_{\G} \left(\mu\mu_{\tt}^{-1} \right) +
  \varphi_{\G} \big(\log^{\vee}_{\G}(\mu\mu_{\tt}^{-1}) \big) \xi \,.\label{eq:unfolding_1}
\end{align}
By computing the expectation and covariance of \eqref{eq:unfolding_1}, we obtain a reparametrized distribution, $\xi_{\tt}
\sim \mathcal{N}(\xi_{\tt}; r_{\tt},\Sigma^{\varphi})$, where
\begin{align}
  \label{eq:reparam}
  r_{\tt}& = \log^{\vee}_{\G} \left(\mu\mu_{\tt}^{-1} \right)\\
  \label{eq:Pr}
  \Sigma^{\varphi} &=
  \varphi_{\G}(r_{\tt}) \Sigma \varphi_{\G}^{T}(r_{\tt}). 
\end{align}
This pdf is equal to the one obtained through the change of coordinates in \eqref{eq:pdf_change}.
%In particular, following the terminology from \cite{Bourmaud2015a},
%the choice of $\mu_t$ in the procedure of changing the coordinates \eqref{eq:pdf_change} 
%is similar to %unfolding a CGD around the chosen value $\mu_t$ hence representing it in the $\mu_t$'s .
%This can be considered as 
%reparameterization of some component respecting wrt the algebra space of $\mu_t$. 
%Thus we can map a CGD component to the algebra of any $\mu \in G$.
%However, by increasing the displacement $r_j$ the accuracy of \eqref{eq:Pr}, via \eqref{eq:bch2}, decreases.
However, obtaining this result by using the procedure of coordinates change through a pdf is important from the
perspective of KL divergence evaluation.
An illustration of unfolding a component $j$ around $\mu_i$, using \eqref{eq:reparam} and \eqref{eq:Pr}, is given
in Fig.~\ref{fig:illustration}.
%\jc{Algebra space tak-tak uveden?}
}
%\end{remark}

The KL divergence between two CGDs 
$\mathcal{G}_i = \mathcal{G}(\mu_i, \Sigma_i)$ and 
$\mathcal{G}_j = \mathcal{G}(\mu_j, \Sigma_j)$
can now be evaluated as
\begin{align}\nonumber %\label{eq:kl_div}
 \mathcal{D}_{\mathrm{KL}}(\mathcal{G}_i,\mathcal{G}_j)
 & \approx \int_{\R^p} p_i(\xi) \log \bigg( \dfrac{p_i(\xi)}{p_j(\xi)} \bigg) \dd \xi =
  D_{\mathrm{KL}}(\mathcal{N}_i,\mathcal{N}_j) \,, \\
%\end{align}
%where 
%\begin{align}
p_i(\xi) \sim \mathcal{N}_i & = \mathcal{N}(\xi; r_{i},\Sigma^{\varphi}_i) \, , \  
	r_{i} = \log^{\wedge}_{\G}(\mu_i \mu_{\tt}^{-1}) \,,\\ \nonumber
p_j(\xi) \sim \mathcal{N}_j & = \mathcal{N}(\xi; r_{j},\Sigma^{\varphi}_j) \, , \
	r_{j} = \log^{\wedge}_{\G}(\mu_j \mu_{\tt}^{-1}) \,,
\end{align}
and $\Sigma^{\varphi} = \varphi_{\G}(r)\Sigma \varphi_{\G}^\T(r)$.
By employing the change of coordinates, we can evaluate the KL divergence of two CGDs similarly as in the case of
`classical' Gaussian distributions, but with reparametrized means and covariances.
The KL divergence is then equal to
\begin{align}
  \mathcal{D}_{\mathrm{KL}}(\mathcal{N}_i, \mathcal{N}_j) = 
  & \frac12 \Big( \mathrm{tr} \big({\Sigma^{\varphi}_j}^{-1} \Sigma^{\varphi}_i \big) - K +
  \log_{\R}\frac{|\Sigma^{\varphi}_j|}{|\Sigma^{\varphi}_i|}  \\ \nonumber
  & \hspace{-1cm} + (r_j - r_i)^{\mathrm{T}} (\Sigma^{\varphi}_j)^{-1}(r_j - r_i)  \Big),
\end{align}
where $\mathrm{tr}(\,.\,)$ and $|\,.\,|$ designate matrix trace and determinant, respectively, while $K$ is the mean vector dimension.
Finally, for mixture components it is necessary to use the scaled symmetrized KL divergence \cite{Amari2009}, which also
takes component weights into account 
\begin{align}\label{eq:skl}
  \mathcal{D}_{s\mathrm{KL}}(w_i \mathcal{N}_i, w_j \mathcal{N}_j) = & \frac12 \Big( 
   (w_i-w_j)\log_{\R}\frac{w_i}{w_j} + \\ \nonumber
  & \hspace{-1cm}
  w_i \mathcal{D}_{\mathrm{KL}}(\mathcal{N}_i, \mathcal{N}_j) + 
  w_j \mathcal{D}_{\mathrm{KL}}(\mathcal{N}_j,\mathcal{N}_i)  \Big) \,.
\end{align}

\vspace{-4mm}
\subsection{Component picking algorithm}\label{sec:component_picking}\noindent
Now that we know how to compute a distance measure between two CGD mixture components, we need to choose an appropriate component picking algorithm which will tell us how to screen the whole mixture and which components to pick for merging.
However, with CGD mixtures there is also another momentum.
If we have $N$ components in the mixture with different weights, how should we approach the problem of
measuring distance, i.e., choosing $\mu_t$ for the change of coordinates?
Should all the distances be calculated with respect to the mean of the component with the highest weight or the lowest
weight?
Or should we `reparametrize' each component on a pairwise basis?
In this letter we study the following five scenarios: (i/ii) all components are reparametrized about the mean of the
component with the highest/lowest weight, (iii) the reparametrization about the identity element, and
(iv/v) components are reparametrized on a pairwise basis by choosing the mean of the component pair with the
higher/lower weight.
For analyzing the five scenarios we use 
two common component picking strategies; 
(i) Exhaustive pairwise \cite{Salmond2009},
and (ii) West's \cite{West1993} algorithms.
The Exhaustive pairwise algorithm determines distances between all components and merges the closest pair,
while West's algorithm sorts the components according to their respective
weights, then finds and merges the component most similar to the first one. 

\vspace{-3mm}
\subsection{Merging the components}\label{sec:merging}\noindent
A component merging algorithm for Gaussian components in $\R^p$
%in the context of Bayesian tracking systems in a cluttered environment 
was proposed in \cite{Salmond2009}:
\begin{align*}
  %w^* = \sum_i w_i, \ 
  r^* \hspace{-1mm} = \frac{1}{w^*} \sum_i w_i r_i,\, %\nonumber\\
  \Sigma^* \hspace{-1mm} = \frac{1}{w^*} \sum_i \Big( w_i \big(\Sigma_i + r_ir_i^{\mathrm{T}}\big) \Big) - r^*(r^*)^{\mathrm{T}}
  %\label{eq:merging_formulae}
\end{align*}
where $w^* = \sum_i w_i$,  % $r^* = \frac{1}{w^*} \sum_i w_i r_i$,  
$w_i\mathcal{N}(r_i, \Sigma_i)$ represents the $i$-th component, and $w^*\mathcal{N}(r^*, \Sigma^*)$ is the resulting component. % $w^*$, $\mu^* \in \R^n$ and $P^*$ designate the parameters resulting from the component merging.
Although merging works for an arbitrary number of components, in our case we will always merge two.

%\jc{Hoćemo li potpuno poopćenit stvari za proizvoljan izbor tangencijalnog prostora?}
However, the previous expressions are defined for Gaussians in $\R^p$ and the question arises how to apply the same
approach for CGD mixtures?
We propose to use the same principle as for computing the KL divergence described in
Section~\ref{sec:component_distance_measure}, i.e., the components to be merged need to be first reparametrized about
the tangent space of the same mean, since covariances are only relevant with respect to their own mean. % and then
%\eqref{eq:merging_formulae} can be applied.
%The simplest example is when one component is reparametrized to the algebra of the other (referent) component.
%
%\begin{gather}
%  w^* = w_i + w_j, \qquad r^* = \frac{w_j}{w^*} r \nonumber\\
%  P^{r^*} = \frac{1}{w^*} \left[ w_i P_i + w_j(P^{r} + r r^{\mathrm{T}})\right]  - r^*(r^*)^{\mathrm{T}}.
%  \label{eq:merging_formulae_group}
%\end{gather}
%
%Note that the mean value of the referent component is always zero.
Once we compute the resulting component, $w^*\mathcal{N}(r^*, \Sigma^*)$, we need to map it back to the group $\G$.
Given a lemma from \cite{Bourmaud2015} and following convention \eqref{eq:cgd}, the procedure evaluates to
\begin{equation} \label{eq:remap_reduction}
\begin{split}
\hspace{-2mm}  \mu^* \hspace{-1mm} = \hspace{-0mm} \exp_{\G}^{\wedge} (r^*)\mu_t , \,
  \Sigma^{\Phi^*} \hspace{-2mm} = \hspace{-0mm} \Ad_{\G}^{r^*} \hspace{-1mm} \Phi_{\G}(r^*) \Sigma^* 
    \big( \hspace{-1mm} \Ad_{\G}^{r^*} \hspace{-1mm} \Phi_{\G}(r^*) \big)^{\mathrm{T}},
\end{split}
\end{equation}
where $\Ad_{\G}^{r^*}=\Ad_{\G}(\exp_{\G}^{\wedge} (r^*))$.
We can notice that covariance reparametrization was necessary to make it relevant from the perspective of the tangent
space of the newly computed $\mu^*$.
An illustration of merging and reparametrization \eqref{eq:remap_reduction} of component $j$ with respect to %the the tangent space of the mean 
$\mu_i$ is given in Fig.~\ref{fig:illustration}.

\vspace{-2mm}
\section{Study example - PHD filter on Lie groups}\noindent
MTT is a complex problem consisting of many challenges and PHD filter presents itself as one of the solutions to MTT.
The reason why PHD filter is interesting for the present letter is because one of its implementations is based on Gaussian
mixtures (GM-PHD) \cite{Vo2006}.
%such as process and sensor noise, false alarms and imperfect detection, as well as measurement origin uncertainty, data association and target births and deaths \cite{Mallick2013}.
%while many scenarios deal with applications where target's state arises on Lie groups, e.g., 
%special Euclidean group
%the target pose as $\SE(2)$ or $\SE(3)$.
%For these applications, an estimation approach operating on Lie groups represents an inevitable tool.
Besides Gaussians, other distributions can be used and in our previous work \cite{Markovic2015} we proposed a
mixture approximation of the PHD filter based on the von Mises distribution on the unit circle. 
In this letter, as a study example, we implement a PHD filter tailored for Lie groups (LG-PHD), %or any combination thereof,
based on the mixture of CGDs and the reduction schemes presented in the previous section.
The LG-PHD can be potentially applied in MTT scenarios where the target state is modelled as a pose in $\SE(2)$ or $\SE(3)$

%\subsection{Discussion}\noindent
The PHD filter propagates the \emph{intensity function} $D_{k-1}$, and operates by evaluating two steps---prediction and update.
By assuming $D_{k-1}$ and birth intensity being Gaussian mixtures \cite{Vo2006},
the GM-PHD prediction results with another Gaussian mixture (Prop. $1$ in \cite{Vo2006}).
%relying on the classical KF prediction applied to each mixture component.
%single-target prediction step of the classical Kalman filter (KF) 
%in a component-wise manner (Prop. $1$ in \cite{Vo2006}).
Similarly, if $D_{k-1}$ and birth intensity are given with CGD mixtures,
the LG-PHD prediction results with another CGD mixture,
relying on the LG-EKF prediction applied to each mixture component \cite{Bourmaud2016}.
%

%in \cite{Bourmaud2016} it was shown that 
%
%By following the same assumptions as in \cite{Vo2006}, but for the CGD, i.e., the posterior intensity $D_{k-1}$ and
%birth intensity being CGD mixtures,
%the predicted intensity $D_{k|k-1}$ will also evaluate to a CGD mixture.

The product of two Gaussians evaluates to a scaled Gaussian, hence
the update step of GM-PHD can be calculated analytically 
(Prop.~$2$ in \cite{Vo2006}).
In contrary, the product of two CGDs, occurring in LG-PHD update, cannot be evaluated directly.
%which shows up in LG-PHD update in the component-wise manner.
%hence we need to alleviate this problem in LG-PHD update step.
Hence, we apply approximations following the same train of thought as in LG-EKF prediction \cite{Bourmaud2016}
where 
given posterior $p(X_{k-1} | Z_{1:k-1})$ and motion model $p(X_k | X_{k-1})$,
it approximates the joint distribution $p(X_k, X_{k-1} | Z_{1:k-1})$,
and then marginalizes obtaining $p(X_k | Z_{1:k-1})$.
%whence the mean and covariance can be obtained 
Similarly,  
given $p(X_k | Z_{1:k-1})$ and likelihood $p(Z_k | X_{k})$,
we approximate the joint distribution $p(X_k, Z_k | Z_{1:k-1})$,
and then marginalize obtaining $p(X_k | Z_k)$.
%The rest of the LG-PHD derivation essentially follows that of GM-PHD.
Final LG-PHD formulae are nearly identical to GM-PHD, except for Jacobian matrices.

%can be approximated by a CGD.
%This approximation follows the same train of thought as LG-EKF prediction,
%which consists of evaluation of the joint distribution and marginalization by way of Gauss-Newton %minimization \cite{Bourmaud2015a},
%In case of the GM-PHD, 
%given that product of 
%the update step is calculated analytically,
%relying on  $2$ in \cite{Vo2006}), 
%since the
%product of Gaussians on a component basis, evaluates to an updated Gaussian scaled by
%the innovation Gaussian \cite{Vo2006}.
%However, for the LG-EKF case, this product is not evaluated directly, but approximated by a posterior CGD whose mean and covariance are obtained by Gauss-Newton minimization \cite{Bourmaud2016}.
%Nevertheless, by following the same train of thought as in \cite{Bourmaud2015a}, 
%for the case of the CGD similar result is obtained as for the Gaussian distribution; but,
%Naturally, the state is now evolving on Lie groups,
%and the corrected intensity $D_{k}$ is approximated with a CGD mixture.
%
%and more involved measurement equation Jacobian.
%Following the same assumptions for the update step as in \cite{Vo2006}, but for the CGD, the corrected intensity
%$D_{k}$ also evaluates to a CGD mixture.

\vspace{-3mm}
\subsection{Experiments}
\begin{figure}[!t]
\vspace{-0mm}
\centering
  %\tikzsetnextfilename{traj_example}
  %\input{figures/traj_example_002.tex}
  \includegraphics[scale=1]{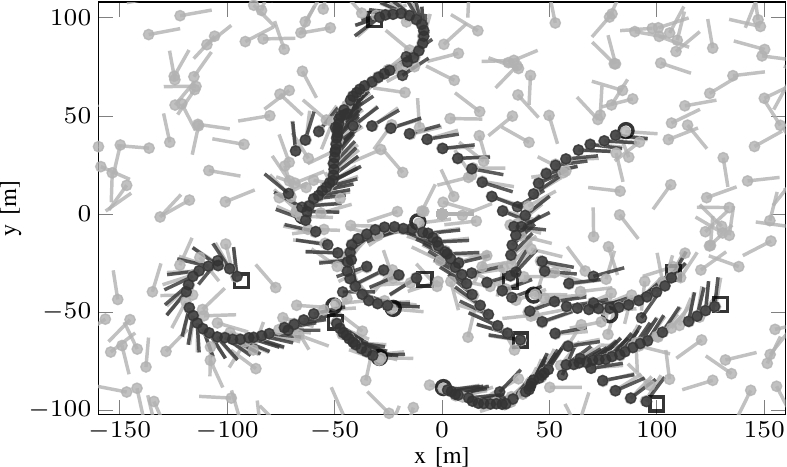}
  \vspace{-2mm}
\caption{An example of a multitarget tracking scenario, involving $10$ objects, out of which $5$ appeared at the beginning, and $5$ more were born during the $100$ steps long sequence (gray arrows--measurements including false alarms, 
black arrows--estimated states, 
black circles--true object birth place, 
black square--true object death place).}
\label{fig:example}
\vspace{-2mm}
\end{figure}

\noindent
In order to validate the performance of the proposed LG-PHD filter, and compare different reduction approaches that are applied after  update steps, we devised appropriate Monte Carlo simulation scenarios.
We applied two component picking strategies, namely the West's algorithm and the pairwise component picking algorithm.
For each we applied the reparametrization approaches as discussed in 
Section~\ref{sec:component_picking}, 
including the mapping to tangent space of 
(i) pairwise larger component $\mathcal{T}_{\text{L}}$, 
(ii) pairwise smaller component $\mathcal{T}_{\text{S}}$, 
(iii) identity element $\mathcal{T}_{\text{Id}}$, 
(iv) largest component $\mathcal{T}_{\text{Max}}$,
(v) smallest component $\mathcal{T}_{\text{Min}}$ 
(West's algorithm always merges the smallest component hence (ii) and (v) are the same).
We generated $100$ examples of an MTT scenario and compared the performance of the approaches.
The initial number of targets in the scene was a random integer $N_{0|0} \in [5, 7]$, while the probability of survival was $p_S = 0.975$ and birth rate was $\lambda_b = 0.25$.
All measurements were corrupted with white noise variance $\sigma_{xy}^2 = 0.5^2\,\mathrm{m}^2$ in distance and $\sigma_{\phi} = 0.1$\,rad/s in orientation, while clutter was governed by the Poisson distribution with intensity $\lambda_Z = 5$.
The state $X = (X^{\text{pos}},X^{\text{vel}}) \in \SE(2) \times \R^3$
contains position and velocity components.
Here we apply the
constant velocity motion model \cite{Cesic2016a} given as 
\begin{align}
&f(X_{k-1}) =
  X_{k-1}
  \exp^{\wedge}_{\G}
  \begin{bmatrix}
    T X^{\text{vel}}_{k-1}\\
    \0
  \end{bmatrix}\,.
\end{align}
We derive the pertaining Jacobian
\begin{align}
&F_{k-1} \hspace{-1mm}=\hspace{-1mm} -\frac{\dd}{\dd s} \hspace{-1mm}
  \left. \bigg(
  \hspace{-1mm} \log^{\vee}_{\G} \Big( f(\mu_{k-1}) f(\exp^{\wedge}_{\G}(s) \mu_{k-1})^{-1} \Big)
  \hspace{-1mm}
  \bigg) \right |_{s=0}\\
& \hspace{-0mm} = \hspace{-1mm}
 \begin{bmatrix}
  \I & T \Phi_{\SE(2)} \bigg( T \Ad_{\SE(2)}\big(\mu_{k-1}^{\text{pos}}\big) \mu^{\text{vel}}_{k-1} \bigg)
  \Ad_{\SE(2)}\big(\mu_{k-1}^{\text{pos}}\big) \\\nonumber
  \0 & \I
 \end{bmatrix} \hspace{-1mm},
\end{align}
where $\mu_{k-1} = (\mu^{\text{pos}}_{k-1},\mu^{\text{vel}}_{k-1}) \in \SE(2) \times \R^3$ 
is the mean value, and $T$ is discretization time.
The probability of measurement detection was $p_D = 0.975$ and the measurements were arising as $\SE(2)$, hence $h(X_k) =
X_k^{\text{pos}} $ and the measurement Jacobian was 
$
H_k = 
\begin{bmatrix}
  \I & \0
\end{bmatrix}
$.
%Complete derivation of the Jacobians is given in the \emph{Supplementary material}.
For illustration purposes, an example of a multitarget scenario with tracking in total $10$ targets on $\SE(2)$ is given in
Fig.~\ref{fig:example} together with LG-PHD results.

%For comparison, 
As a performance metric we used the optimal subpattern asignement (OSPA) metric \cite{Schuhmacher2008}.
In Table~\ref{tab:statistics} we present the results where for each of the $100$ multitarget trajectories
the cumulative OSPA $\mathcal{D}_{\text{t}}$, and its localization component $\mathcal{D}_{\text{d}}$ and
cardinality component $\mathcal{D}_{\text{c}}$ were calculated.
\begin{table}[!t]
\vspace{-0mm}
\caption{Average OSPA over $100$ multitarget scenarios.}\label{tab:statistics}
\vspace{-3mm}
\begin{center}
\begin{tabular}{m{0.375cm} m{0.375cm} m{0.375cm} m{0.375cm} m{0.375cm} m{0.375cm} m{0.05cm} m{0.375cm} m{0.375cm} m{0.375cm} m{0.5cm}}
\toprule
       & \multicolumn{5}{c}{Exhaustive pairwise} & & \multicolumn{4}{c}{West} \\\cmidrule{2-6}\cmidrule{8-11}
	   & $\mathcal{T}_{\text{L}}$ & $\mathcal{T}_{\text{S}}$ & $\mathcal{T}_{\text{Id}}$ & $\mathcal{T}_{\text{Max}}$ & $\mathcal{T}_{\text{Min}}$ & & $\mathcal{T}_{\text{L}}$ & $\mathcal{T}_{\text{S}}$ & $\mathcal{T}_{\text{Id}}$ & $\mathcal{T}_{\text{Max}}$ \\ \midrule
$\mathcal{D}_{\text{t}}$ & 
\textbf{2.445} & 2.515 & 2.764 & 2.912 & 3.082 & & \textbf{1.910} & 1.924 & 2.060 & 2.125 \\ \midrule
$\mathcal{D}_{\text{d}}$ & 
\textbf{2.100} & 2.163 & 2.419 & 2.558 & 2.695 & & \textbf{1.415} & 1.420 & 1.537 & 1.605 \\ \midrule
$\mathcal{D}_{\text{c}}$ & 
\textbf{0.594} & 0.613 & 0.627 & 0.653 & 0.745 & & \textbf{0.737} & 0.746 & 0.797 & 0.792 \\ % \bottomrule
%\toprule
%       & \multicolumn{5}{c}{West} \\\cmidrule{2-6}
%	   & $\mathcal{T}_{\text{L}}$ & $\mathcal{T}_{\text{S}}$ & $\mathcal{T}_{\text{Id}}$ & $\mathcal{T}_{\text{Max}}$ & $\mathcal{T}_{\text{Min}}$ \\ \midrule
%$\mathcal{D}_{\text{t}}$ & 88.23 & 88.46 & 90.05 & 89.25 & 88.46 \\ \midrule
%$\mathcal{D}_{\text{d}}$ & 67.43 & 67.45 & 69.25 & 68.35 & 67.45 \\ \midrule
%$\mathcal{D}_{\text{c}}$ & 25.53 & 25.94 & 25.79 & 26.03 & 25.94 \\
\bottomrule
\end{tabular}
\end{center}
\vspace{-5mm}
\end{table}%
For both Exhaustive pairwise and West's picking strategies, 
relying on mapping to the tangent space of pairwise larger components $\mathcal{T}_{\text{L}}$ generally outperformed the other approaches.

\vspace{-2mm}
\section{Conclusion}
\label{sec:con}\noindent
In this letter we have studied the problem of mixture reduction on matrix Lie groups.
We have particularly dealt with the manipulation of CGD components to compute the KL divergence, pick and
merge the mixture components.
As a study example, we implemented the LG-PHD filter, a mixture approximation
of the PHD filter tailored for MTT with states evolving on matrix Lie groups.
Using the OSPA metric we analyzed the performance of the LG-PHD filter with respect to mixture component number
reduction.

%\appendices
%\section{}\label{app:omega}

%\section*{Acknowledgment}

\ifCLASSOPTIONcaptionsoff
  \newpage
\fi

%\newpage
\balance
\bibliographystyle{IEEEtran}
\bibliography{bibliography/bibliography,bibliography/library}

\begin{thebibliography}{10}
\providecommand{\url}[1]{#1}
\csname url@samestyle\endcsname
\providecommand{\newblock}{\relax}
\providecommand{\bibinfo}[2]{#2}
\providecommand{\BIBentrySTDinterwordspacing}{\spaceskip=0pt\relax}
\providecommand{\BIBentryALTinterwordstretchfactor}{4}
\providecommand{\BIBentryALTinterwordspacing}{\spaceskip=\fontdimen2\font plus
\BIBentryALTinterwordstretchfactor\fontdimen3\font minus
  \fontdimen4\font\relax}
\providecommand{\BIBforeignlanguage}[2]{{%
\expandafter\ifx\csname l@#1\endcsname\relax
\typeout{** WARNING: IEEEtran.bst: No hyphenation pattern has been}%
\typeout{** loaded for the language `#1'. Using the pattern for}%
\typeout{** the default language instead.}%
\else
\language=\csname l@#1\endcsname
\fi
#2}}
\providecommand{\BIBdecl}{\relax}
\BIBdecl

\bibitem{Alspach1972}
D.~Alspach and H.~Sorenson, ``Nonlinear bayesian estimation using gaussian sum
  approximations,'' \emph{IEEE Transactions on Automatic Control}, vol.~17,
  no.~4, pp. 439--448, 1972.

\bibitem{Chen2000}
R.~Chen and J.~S. Liu, ``{Mixture Kalman filters},'' \emph{Journal of the Royal
  Statistical Society: Series B (Statistical Methodology)}, vol.~62, no.~3, pp.
  493--508, 2000.

\bibitem{Blackman1986}
S.~S. {Blackman}, \emph{{Multiple-target tracking with radar applications}},
  {Gilmour}, Ed., 1986.

\bibitem{Stauffer2000}
C.~Stauffer and W.~E.~L. Grimson, ``Learning patterns of activity using
  real-time tracking,'' \emph{IEEE Transactions on Pattern Analysis and Machine
  Intelligence}, vol.~22, no.~8, pp. 747--757, Aug 2000.

\bibitem{Goldberger2005}
J.~Goldberger and H.~Aronowitz, ``{A distance measure between GMMs based on the
  unscented transform and its application to speaker recognition},'' in
  \emph{Proceedings of Interspeech}, 2005, pp. 1985--1989.

\bibitem{Nikseresht2008}
A.~Nikseresht and M.~Gelgon, ``Gossip-based computation of a gaussian mixture
  model for distributed multimedia indexing,'' \emph{IEEE Transactions on
  Multimedia}, vol.~10, no.~3, pp. 385--392, 2008.

\bibitem{Salmond1989}
D.~Salmond, ``{Mixture reduction algorithms for target tracking},'' in
  \emph{IEEE Colloquium on State Estimation in Aerospace and State Estimation},
  1989, pp. 1--4.

\bibitem{Runnalls2007}
A.~Runnalls, ``{Kullback-Leibler approach to Gaussian mixture reduction},''
  \emph{IEEE Transactions on Aerospace and Electronic Systems}, vol.~43, no.~3,
  pp. 989--999, 2007.

\bibitem{Bukal2014a}
M.~Bukal, I.~Markovi{\'{c}}, and I.~Petrovi{\'{c}}, ``{Composite distance based
  approach to von Mises mixture reduction},'' \emph{Information Fusion},
  vol.~20, no.~1, pp. 136--145, 2014.

\bibitem{Vo2006}
B.-N. Vo and W.-K. Ma, ``{The Gaussian mixture probability hypothesis density
  filter},'' \emph{IEEE Transactions on Signal Processing}, vol.~54, no.~11,
  pp. 4091--4104, 2006.

\bibitem{Vo2007}
B.-T. Vo, B.-N. Vo, and A.~Cantoni, ``{Analytic implementations of the
  cardinalized probability hypothesis density filter},'' \emph{IEEE
  Transactions on Signal Processing}, vol.~55, no.~7, pp. 3553--3567, 2007.

\bibitem{Lui2012}
Y.~M. Lui, ``{Advances in matrix manifolds for computer vision},'' \emph{Image
  and Vision Computing}, vol.~30, no. 6-7, pp. 380--388, 2012.

\bibitem{Loianno2016}
G.~Loianno, M.~Watterson, and V.~Kumar, ``{Visual inertial odometry for
  quadrotors on SE(3)},'' in \emph{IEEE International Conference on Robotics
  and Automation (ICRA)}, 2016, pp. 1544--1551.

\bibitem{Cesic2016b}
J.~{\'{C}}esi{\'{c}}, V.~Joukov, I.~Petrovi{\'{c}}, and D.~Kuli{\'{c}}, ``{Full
  body human motion estimation on Lie groups using 3D marker position
  measurements},'' in \emph{IEEE-RAS International Conference on Humanoid
  Robots (Humanoids)}, 2016, pp. 826--833.

\bibitem{Barrau2015}
A.~Barrau and S.~Bonnabel, ``{Intrinsic filtering on Lie groups with
  applications to attitude estimation},'' \emph{IEEE Transactions on Automatic
  Control}, vol.~60, no.~2, pp. 436--449, 2015.

\bibitem{Long2012}
A.~W. Long, K.~C. Wolfe, M.~J. Mashner, and G.~S. Chirikjian, ``{The banana
  distribution is Gaussian: A localization study with exponential
  coordinates},'' in \emph{Proceedings of Robotics: Science and Systems (RSS)},
  Sydney, Australia, 2012.

\bibitem{Barfoot2014}
T.~D. Barfoot and P.~T. Furgale, ``{Associating uncertainty with
  three-dimensional poses for use in estimation problems},'' \emph{IEEE
  Transactions on Robotics}, vol.~30, no.~3, pp. 679--693, Jun. 2014.

\bibitem{Zhang2016}
C.~Zhang, A.~Taghvaei, and P.~G. Mehta, ``{Feedback particle filter on matrix
  Lie groups},'' in \emph{Proceedings of the American Control Conference},
  2016, pp. 2723--2728.

\bibitem{Bourmaud2013}
G.~Bourmaud, R.~M{\'{e}}gret, A.~Giremus, and Y.~Berthoumieu, ``{Discrete
  extended Kalman filter on Lie groups},'' in \emph{European Signal Processing
  Conference (EUSIPCO)}, 2013, pp. 1--5.

\bibitem{Bourmaud2015}
G.~Bourmaud, R.~M{\'{e}}gret, M.~Arnaudon, and A.~Giremus,
  ``{Continuous-discrete extended Kalman filter on matrix Lie groups using
  concentrated Gaussian distributions},'' \emph{Journal of Mathematical Imaging
  and Vision}, vol.~51, no.~1, pp. 209--228, 2015.

\bibitem{Wolfe2011}
K.~C. Wolfe, M.~Mashner, and G.~S. Chirikjian, ``{Bayesian fusion on Lie
  groups},'' \emph{Journal of Algebraic Statistics}, vol.~2, no.~1, pp. 75--97,
  2011.

\bibitem{Chirikjian2012}
G.~S. Chirikjian, \emph{{Stochastic Models, Information Theory, and Lie
  Groups}}.\hskip 1em plus 0.5em minus 0.4em\relax Birkh{\"{a}}user, 2012,
  vol.~2.

\bibitem{Bourmaud2016}
G.~Bourmaud, R.~M{\'{e}}gret, A.~Giremus, and Y.~Berthoumieu, ``{From intrinsic
  optimization to iterated extended Kalman filtering on Lie groups},''
  \emph{Journal of Mathematical Imaging and Vision}, vol.~55, no.~3, pp.
  284--303, 2016.

\bibitem{Bourmaud2015a}
G.~Bourmaud, ``{Estimation de param{\`{e}}tres {\'{e}}voluant sur des groupes
  de Lie : Application {\`{a}} la cartographie et {\`{a}} la localisation d'une
  cam{\'{e}}ra monoculaire},'' Ph.D. dissertation, University of Bordeaux,
  2015.

\bibitem{Miller1972}
W.~Miller, \emph{{Symmetry Groups and their Applications}}, 1972.

\bibitem{Kullback1997}
S.~Kullback, \emph{{Information Theory and Statistics}}.\hskip 1em plus 0.5em
  minus 0.4em\relax New York: Dover Publications, 1997.

\bibitem{Amari2009}
S.~Amari, ``{Alpha-divergence is unique, belonging to both f-divergence and
  Bregman divergence classes},'' \emph{IEEE Transactions on Information
  Theory}, vol.~55, no.~11, pp. 4925--4931, 2009.

\bibitem{Salmond2009}
D.~J. Salmond, ``{Mixture reduction algorithms for point and extended object
  tracking in clutter},'' \emph{IEEE Transactions on Aerospace and Electronic
  Systems}, vol.~45, no.~2, pp. 667--686, 2009.

\bibitem{West1993}
M.~West, ``{Approximating posterior distributions by mixtures},'' \emph{Journal
  of Royal Statistical Society, Series B}, vol.~55, no.~2, pp. 409--442, 1993.

\bibitem{Markovic2015}
I.~Markovi{\'{c}}, J.~{\'{C}}esi{\'{c}}, and I.~Petrovi{\'{c}}, ``{Von Mises
  Mixture PHD Filter},'' \emph{IEEE Signal Processing Letters}, vol.~20,
  no.~12, pp. 2229--2233, 2015.

\bibitem{Cesic2016a}
J.~{\'{C}}esi{\'{c}}, I.~Markovi{\'{c}}, and I.~Petrovi{\'{c}}, ``{Moving
  object tracking employing rigid body motion on matrix Lie groups},'' in
  \emph{International Conference on Information Fusion (FUSION)}, 2016, p.~7.

\bibitem{Schuhmacher2008}
D.~Schuhmacher, B.~T. Vo, and B.~N. Vo, ``{A consistent metric for performance
  evaluation of multi-object filters},'' \emph{IEEE Transactions on Signal
  Processing}, vol.~56, pp. 3447--3457, 2008.

\end{thebibliography}

\end{document}